\documentclass[superscriptaddress,twocolumn]{revtex4-1}

\usepackage{amsfonts}
\usepackage{amsmath}    
\usepackage[normalem]{ulem}
\usepackage{bm}
\usepackage{graphicx}
\usepackage{placeins}

\newcommand{\ket}[1]{\vert#1\rangle}

\def\opone{\leavevmode\hbox{\small1\kern-3.8pt\normalsize1}}
\usepackage{color}

\begin{document}

\title{A multiplexed light-matter interface for fibre-based quantum networks}

\author{Erhan Saglamyurek}
\affiliation{Institute for Quantum Science and Technology, and Department of Physics \& Astronomy, University of Calgary, 2500 University Drive NW, Calgary, Alberta T2N 1N4, Canada}

\author{Marcel.li {Grimau Puigibert}}
\affiliation{Institute for Quantum Science and Technology, and Department of Physics \& Astronomy, University of Calgary, 2500 University Drive NW, Calgary, Alberta T2N 1N4, Canada}

\author{Qiang Zhou}
\affiliation{Institute for Quantum Science and Technology, and Department of Physics \& Astronomy, University of Calgary, 2500 University Drive NW, Calgary, Alberta T2N 1N4, Canada}

\author{Lambert Giner}
\altaffiliation{Present address: Department of Physics, University of Ottawa, 150 Louis Pasteur, Ottawa, Ontario, K1N 6N5 Canada}
\affiliation{Institute for Quantum Science and Technology, and Department of Physics \& Astronomy, University of Calgary, 2500 University Drive NW, Calgary, Alberta T2N 1N4, Canada}

\author{Francesco Marsili}
\affiliation{Jet Propulsion Laboratory, California Institute of Technology, 4800 Oak Grove Drive, Pasadena, California 91109, USA}

\author{Varun B. Verma}
\affiliation{National Institute of Standards and Technology, Boulder, Colorado 80305, USA}

\author{Sae Woo Nam}
\affiliation{National Institute of Standards and Technology, Boulder, Colorado 80305, USA}

\author{Lee Oesterling}
\affiliation{Battelle, 505 King Ave, Columbus, OH 43201, USA}

\author{David Nippa}
\affiliation{Battelle, 505 King Ave, Columbus, OH 43201, USA}

\author{Daniel Oblak}
\affiliation{Institute for Quantum Science and Technology, and Department of Physics \& Astronomy, University of Calgary, 2500 University Drive NW, Calgary, Alberta T2N 1N4, Canada}

\author{Wolfgang Tittel}
\affiliation{Institute for Quantum Science and Technology, and Department of Physics \& Astronomy, University of Calgary, 2500 University Drive NW, Calgary, Alberta T2N 1N4, Canada}

\maketitle

\textbf{Processing and distributing quantum information using photons through fibre-optic or free-space links is essential for building future quantum networks. The scalability needed for such networks can be achieved by employing photonic quantum states that are multiplexed into time and/or frequency, and light-matter interfaces that are able to store and process such states with large time-bandwidth product and multimode capacities. Despite important progress in developing such devices, the demonstration of these capabilities using non-classical light remains challenging. Employing the atomic frequency comb quantum memory protocol in a cryogenically cooled erbium-doped optical fibre, we report the quantum storage of heralded single photons at a telecom-wavelength (1.53 $\mu$m) with a time-bandwidth product approaching 800. Furthermore we demonstrate frequency-multimode storage as well as memory-based spectral-temporal photon manipulation. Notably, our demonstrations rely on fully integrated quantum technologies operating at telecommunication wavelengths, i.e. a fibre-pigtailed nonlinear waveguide for the generation of heralded single photons, an erbium-doped fibre for photon storage and manipulation, and fibre interfaced superconducting nanowire devices for efficient single photon detection. With improved storage efficiency, our light-matter interface may become a useful tool in future quantum networks.} 

Multiplexing, particularly in the form of wavelength division multiplexing, is key for achieving high data rates in modern fibre-optic communication networks. The realization of  scalable quantum information processing demands adapting this concept, if possible using components that are compatible with the existing telecom infrastructure. One of the challenges to achieve this goal is to develop integrated light-matter interfaces that allow storing and processing multiplexed photonic quantum information. In addition to efficient operation, ease of integration and feed-forward controlled recall, the suitability of such interfaces depends on the storage time for a given acceptance bandwidth, i.e. the interface's time-bandwidth product, as well as  its multimode storage and processing capacities. It is important to note that while the multimode capacity of a light-matter interface determines the maximum number of simultaneously storable and processable photonic modes, the time-bandwidth product only sets an upper bound to the multimode capacity for temporally and/or spectrally multiplexed photons.

\begin{figure*} [t!]
\begin{center}
\includegraphics[width=1\textwidth,angle=0]{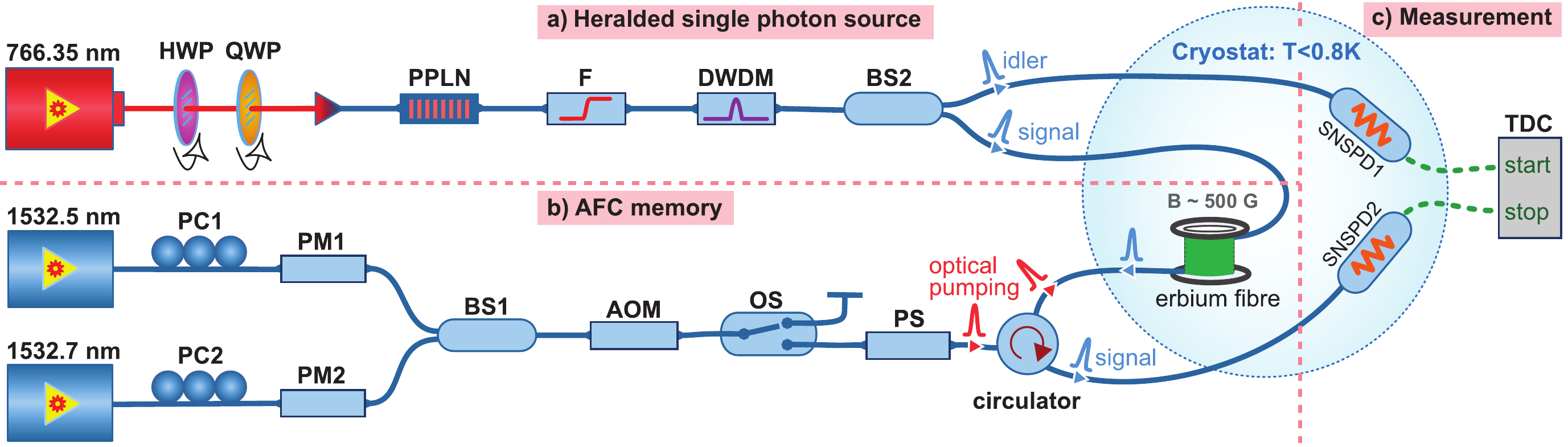}
\caption{\textbf{Experimental setup.} 
\textbf{a) Heralded single photon source.} Narrow linewidth continuous-wave (CW) light at 766.35~nm wavelength and with 100~$\mu$W power is sent to a fibre-pigtailed, periodically-poled lithium niobate (PPLN) waveguide that is heated to 52.8$^{\circ}$C. A quarter-wave-plate (QWP) and a half-wave-plate (HWP) match the polarization of the light to the crystal's C axis to maximize the non-linear interaction. Spontaneous parametric down conversion (SPDC) in the PPLN crystal results in frequency degenerate photon pairs with 40~nm bandwidth, centred at 1532~nm wavelength. The residual pump light at 766.35~nm is suppressed by 50~dB by a filter (F) and the bandwidth of the created photons is filtered down to 50~GHz using a dense-wavelength-division-multiplexer (DWDM). The filtered photon pairs are probabilistically split using a beam-splitter (BS2). The detection of one member (the \textit{idler} photon) heralds the presence of the other (the \textit{signal} photon), which is directed to the input of the AFC memory. 
\textbf{b) Quantum memory.} The quantum memory is based on an erbium-doped fibre that is exposed to a magnetic field of 600~G and cooled to a temperature below 1~K. Light from two independent CW lasers with wavelengths of 1532.5 and 1532.7~nm, respectively, is used to spectrally tailor the inhomogeneously broadened 1532 nm absorption line of erbium through frequency-selective optical pumping into one or several atomic frequency combs. Towards this end, phase-modulators (PM1 and PM2) followed by an acousto-optic modulator (AOM) are used to generate chirped pulses with the required frequency spectrum. The optical pumping light from the two lasers is merged on a beam-splitter (BS1) and enters the erbium fibre from the back via an optical circulator. Polarization controllers (PM1 and PM2) match the polarization to the phase-modulators active axes, and the polarization scrambler (PS) ensures uniform optical pumping of all erbium ions in the fibre \cite{jin2015b}.
\textbf{c) Measurement unit.}
The detection of the heralding photon (\textit{idler}) and subsequently the \textit{signal} photon is performed by two superconducting nano-wire single photon detectors (SNSPD1 and SNSPD2) maintained at the same temperature as the memory. The coincidence analysis of the detection events is performed by a time-to-digital converter (TDC).
}
\label{setup} 
\end{center}
\end{figure*}

Over the past decade there has been significant progress towards the creation of such light-matter interfaces. One promising approach is based on a far off-resonant Raman transfer in warm atomic vapour, for which a time-bandwidth product of 5000 has been reported \cite{england2011a}. However, due to noise arising from (undesired) four-wave mixing, the storage and recall of quantum states of light has proven elusive \cite{michelberger2015a}. This problem can be overcome using different media, e.g., diamonds (with storage in optical phonon modes) and laser-cooled atomic ensembles, for which time-bandwidth products around 22 have been obtained with non-classical light \cite{england2015a,ding2015a}. However, the  multimode operation of a Raman-type memory and hence the utilization of the potentially achievable large time-bandwidth products will remain challenging due to unfavourable scaling of this scheme's multimode capacity with respect to optical depth \cite{nunn2008a}. 

Another promising avenue for a multiplexed light-matter interface is the atomic frequency comb (AFC)-based quantum memory scheme in cryogenically cooled rare-earth ion doped materials \cite{riedmatten2008a,afzelius2009a}. An attractive feature of this approach is that the multimode storage capacity is independent of optical depth; it is solely given by the time-bandwidth product of the storage medium, which can easily go up to several thousands due to the generally large inhomogeneous broadening (enabling large storage bandwidth) and narrow homogeneous linewidth (allowing long storage times) of optical transitions in rare-earth ion doped materials. This aspect has already allowed several important demonstrations, including the simultaneous storage of 64 and 1060 temporal modes by AFCs featuring pre-programmed delays \cite{usmani2010a,bonarota2011a},  5 temporal modes by an AFC with recall on-demand \cite{gundogan2013a}, and 26 spectral modes supplemented with frequency-selective recall \cite{sinclair2014a}, respectively. Despite the importance of these demonstrations, they were restricted to the use of strong or attenuated laser pulses rather than non-classical light, as required in future quantum networks. The only exception is the very recent demonstration of the storage of photons, emitted by a quantum dot, in up to 100 temporal modes \cite{tang2015a}.

In this paper we present a spectrally multiplexed light-matter quantum interface for non-classical light. More precisely, we demonstrate large time-bandwidth-product and multimode storage of heralded single-photons at telecom wavelength by implementing the AFC protocol in an ensemble of erbium ions. As an important feature for future quantum networks, our demonstrations rely on fully integrated quantum technologies, i.e. a fibre-pigtailed LiNbO$_3$ waveguide for the generation of heralded single photons by means of parametric down-conversion, a commercially available, cryogenically-cooled erbium-doped single-mode fibre for photon storage and manipulation, and superconducting nanowire devices for high efficiency single photon detection. 

\subsection*{Experiment}

Our experimental setup, illustrated in Figure \ref{setup}, is composed of an integrated, heralded single photon source, an AFC-based erbium-doped fibre memory, and a measurement unit including two superconducting nanowire single-photon detectors (SNSPDs). 

We generate pairs of energy-time quantum correlated telecom-wavelength photons -- commonly referred-to as \textit{signal} (\textit{s}) and \textit{idler} (\textit{i}) -- by sending pump light from a continuous wave (CW) laser operating at 766 nm wavelength to a periodically poled lithium niobate (PPLN) waveguide, as shown in Fig \ref{setup}a. Spontaneous parametric down conversion (SPDC) based on type-0 phase matching results in the creation of frequency-degenerate photon pairs centred at 1532 nm wavelength and having a bandwidth of about 40~nm. We note that the PPLN waveguide is fibre pigtailed at both input and output faces (i.e. for the pump light as well as the down-converted photons), which makes the source alignment free. After filtering away the remaining pump light, the spectra of the generated photons are filtered down to 50~GHz resulting in a photon-pair generation rate of 0.35~MHz. The ensuing photons are probabilistically separated into two standard telecommunication fibres using a 50/50 fibre-optic beam splitter (BS2).  One member of each split pair is sent into an SNSPD featuring a system detection efficiency of around 70\% (see the Methods for details of the SNSPDs). Its electronic output heralds the other member, which travels through standard telecommunication fibre to our light-matter interface (for more details about the heralded single photon source see the Appendix~1).

\begin{figure}[t!]
\begin{center}
\includegraphics[width=\columnwidth,angle=0]{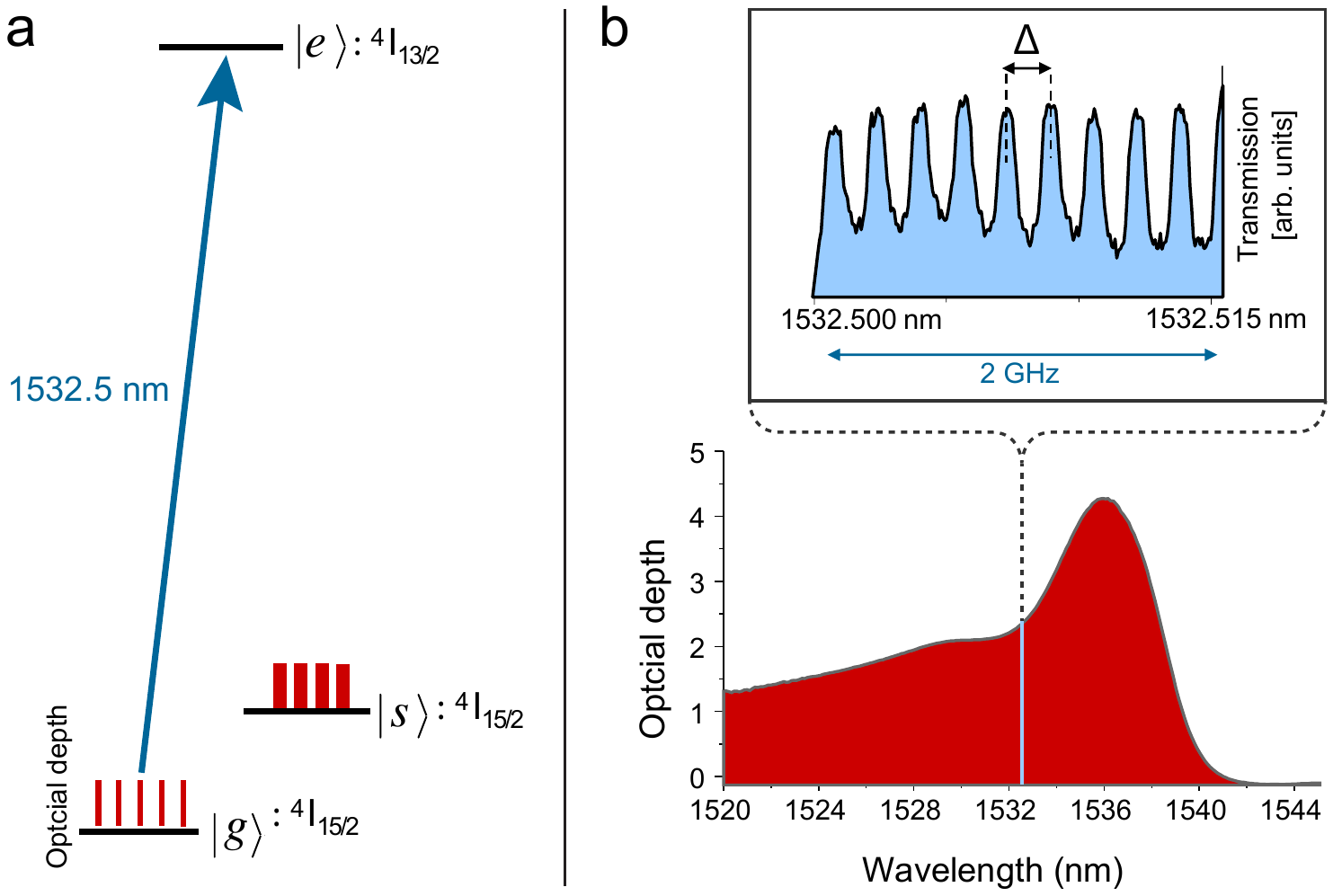}
\caption{\textbf{Quantum memory.}~\textbf{a) Simplified level scheme of Er$^{3+}$ in silica glass.} Frequency-selective optical pumping from the $^4I_{15/2}$ electronic ground state ($\ket{\mathrm{g}}$) via the $^4I_{13/2}$ excited state ($\ket{\mathrm{e}}$) into an auxiliary (spin) state ($\ket{\mathrm{s}}$) allows spectral tailoring.~\textbf{b) Inhomogeneous broadening and AFC structure.} The inhomogeneously broadened optical absorption line of erbium ions in silica fibre at 1~K extends from roughly from 1500 nm to 1540 nm wavelength. A 2 GHz-wide section of a 16 GHz-wide comb at 1532.5~nm with teeth spacing $\Delta$=200~MHz is shown.}      
\label{broadbandstorage}
\end{center}
\end{figure}

To store or manipulate the heralded single photons, we prepare an AFC-based memory in a twenty-meter long erbium-doped silica fibre maintained at a temperature of 0.6-0.8 K and exposed to a magnetic field of 600 G (see Fig.~\ref{setup}b and the Methods for details). The memory relies on spectral tailoring of the inhomogenously broadened, $^4I_{15/2}\leftrightarrow^4I_{13/2}$ transition in erbium into a comb-shaped absorption feature characterized by the teeth spacing, $\Delta$, as shown in Fig.~\ref{broadbandstorage}a. The spectral tailoring is performed by frequency-selective optical pumping of ions into long-lived auxiliary (spin) levels. When an input photon is absorbed by the ions constituting the comb, a collective atomic excitation is created. It is described by:
\begin{equation}
	\label{eq:afcqstate}
	\left| \Psi  \right\rangle =\frac{1}{\sqrt{N}}\sum _{j=1}^{N} c_{j} e^{i2\pi m_j\Delta t} e^{-ikz_j}  \left| g_{1} ,\cdots e_{j} ,\cdots g_{N}  \right\rangle \ ,
\end{equation}
where $N$ is the total number of addressed atoms, $k$ is the optical wave number and ${\left| g_{j}  \right\rangle}$ and ${\left| e_{j}  \right\rangle}$ are the $j$'th atom's ground and excited states, respectively. The detuning of the atom's transition frequency from the photon carrier frequency is given by $m_j\Delta$, while $z_j$ is the position of the atom measured along the propagation direction of the light, and the factor $c_j$ depends on both the resonance frequency and position of the atom.
Due to the periodic nature of the AFC, the atomic excitation is converted back to photonic form and the input photon is re-emitted in the originally encoded state after a storage time given by the inverse of the peak spacing, $t_\mathrm{storage}=1/\Delta$, as shown in Fig.~\ref{broadbandstorage}b. The use of erbium doped fibre is particularly attractive for AFC-based quantum memory because of its polarization insensitive operation at wavelengths within the telecom C-band \cite{jin2015b}, its ability to store photonic entanglement in combination of ease of integration with standard fibre infrastructure \cite{saglamyurek2015a}, and its large usable inhomogeneously broadened absorption line \cite{saglamyurek2015b}, which allows for multimode storage and manipulation of photons with large bandwidth, as detailed below.

\begin{figure}[t!]
\begin{center}
\includegraphics[width=\columnwidth,angle=0]{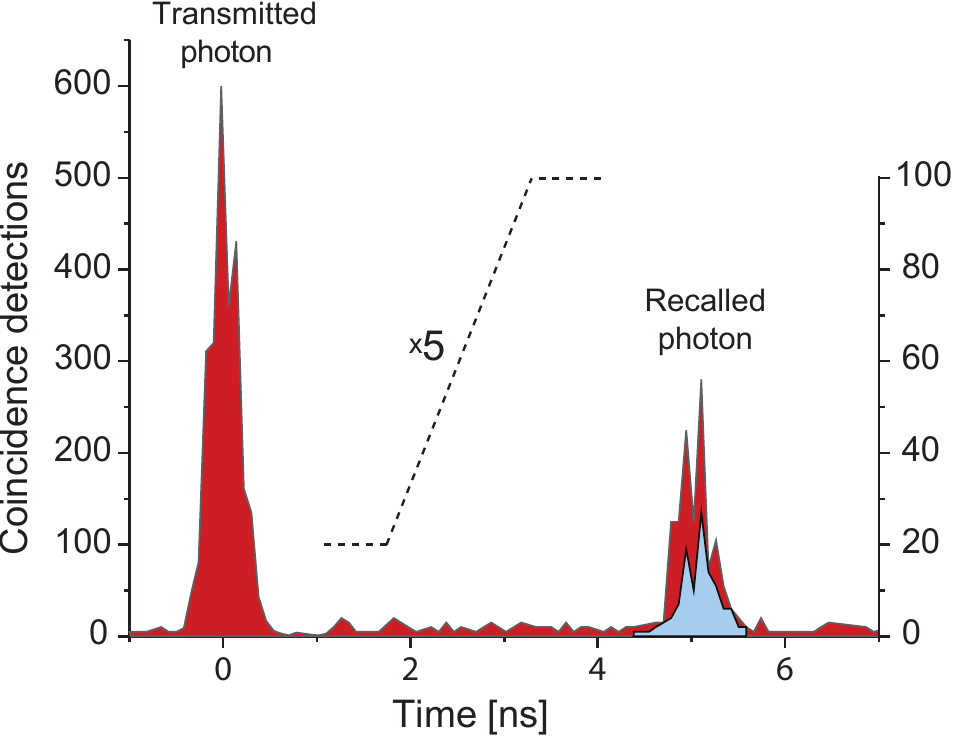}
\caption{\textbf{Reversible mapping of broadband, heralded single photons.} Heralded telecom-wavelength photons, centred at 1532.7 nm wavelength and having a bandwidth of 50 GHz, are mapped onto the AFC memory and recalled after time $t_\mathrm{storage}=\frac{1}{\Delta}=5$~ns. The histogram shows time-resolved coincidence detections collected over 3 minutes. Due to non-unit absorption probability by the AFC as well as bandwidth mismatch between the spectra of the photons and the AFC, a significant portion of the input photons is directly transmitted and detected at zero delay. The light blue highlighted section bounded by a dotted line in the trace of the recalled photon (counts scaled by a factor of five) corresponds to the measurement in which the total AFC bandwidth was 8 GHz; the red section corresponds to a measurement using a 16 GHz wide AFC.}      
\label{broadbandstorage-2}
\end{center}
\end{figure}

Finally, we detect the heralded photons after storage/manipulation using a second SNSPD, featuring similar performance as that used to detect the heralding photon. All detection signals are sent to a time-to-digital converter (TDC) to perform time-resolved coincidence measurements. This allows us to calculate the cross-correlation function 
 \begin{equation}
 \label{g2}
  g_{si}^{(2)}=\frac{R_{si}}{R_{i}R_{s}},
	\end{equation}
where $R_{si}$ is the rate of coincidence detections, and  $R_{s}$ and $R_{i}$ are the single detection count-rate for \textit{signal} and \textit{idler} photons, respectively. 
A classical field satisfies the Cauchy-Schwarz inequality $\left[g_{si}^{(2)}\right]^2 \leq g_{s}^{(2)}g_{i}^{(2)}$ where $g_{s}^{(2)}$ and $g_{i}^{(2)}$ are second-order auto-correlation functions for \textit{signal} and \textit{idler} modes. For photons derived from an SPDC process, the second-order auto-correlation is bounded by $1\leq g_{s,i}^{(2)}\leq2$ \cite{tapster1998a}. Consequently, measuring a cross-correlation $g_{si}^{(2)}$ greater than 2 violates the inequality and thus verifies the presence of quantum correlations between the members of the photon pairs. We characterize our source for different SPDC pump powers (shown in Appendix Fig.~A1), finding that the cross-correlation function exceeds 1000 for all powers. Since $g_{si}^{(2)}\gg2$ implies that the heralded autocorrelation function of the stored signal photons is $\ll 1$ \cite{bashkansky2014a}, we denote these as single photons. The details of how $g_{si}^{(2)}$ is obtained from the coincidence counts are given in the Methods.

\subsection*{Results}

\begin{figure}[t!]
\begin{center}
\includegraphics[width=1\columnwidth,angle=0]{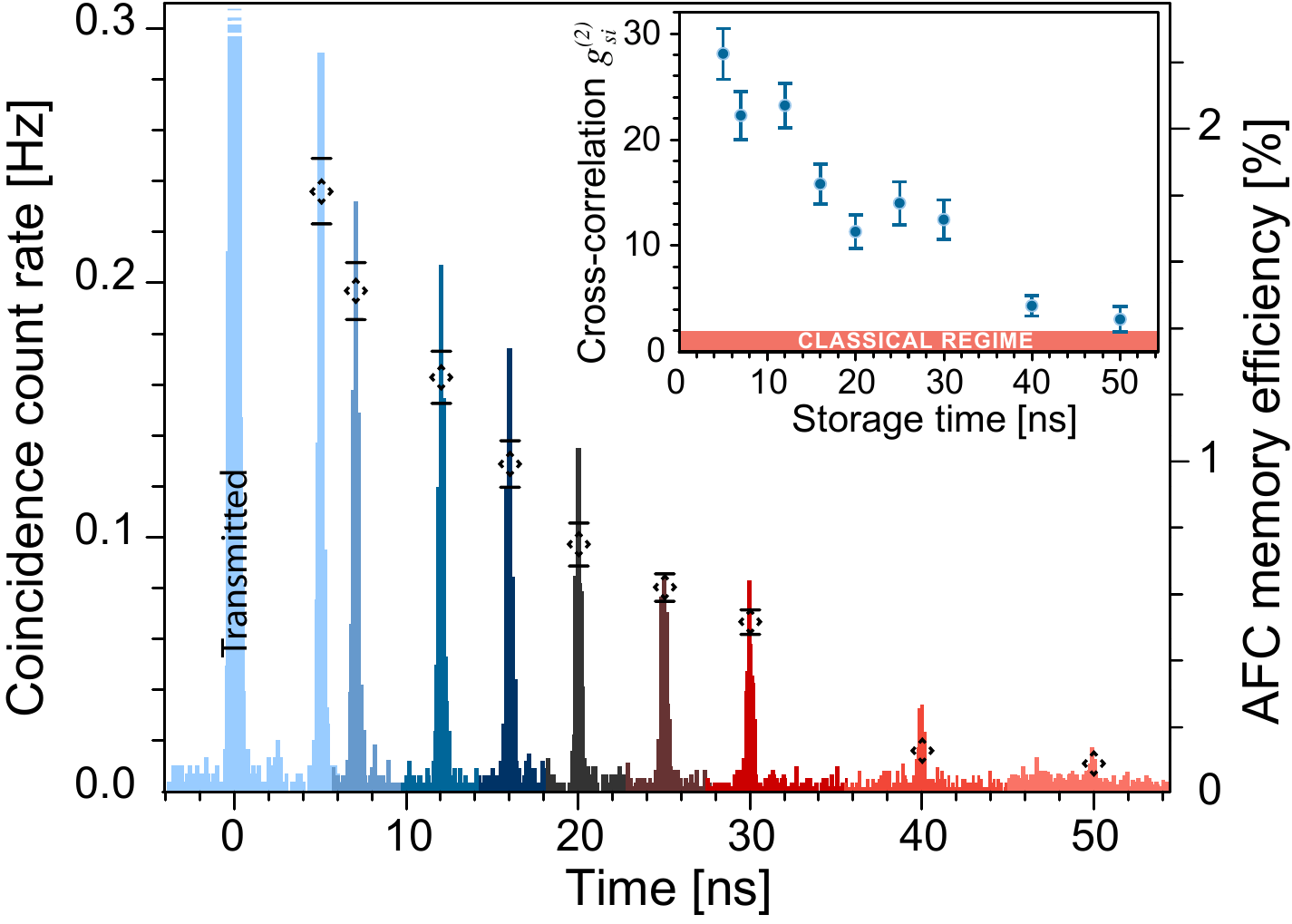}
\caption{\textbf{Quantum storage with large time-bandwidth product.} 16 GHz wide AFC regions with teeth spacing ranging from 200 MHz to 20 MHz are subsequently programmed to store heralded single photons for 5 ns to 50 ns. The histograms show measured coincidence rates for recalled photons for each storage time, and the dashed diamonds depict the corresponding memory efficiencies (see Appendix~2). 
Note that in this figure and henceforth the transmitted pulse at $t=0$ exceeds the vertical scale and thus is capped at the top. Experimentally obtained $g_{si}^{(2)}$ for each storage time are shown in the inset. The measurement time varied between 5 and 15 minutes.}
\label{storagetime}
\end{center}
\end{figure}

First, we investigate the storage of broadband heralded single-photons in an AFC memory -- prepared using a single optical pumping laser at 1532.5~nm -- with a total bandwidth of 8 GHz and 200 MHz tooth spacing (corresponding to 5~ns storage time), as shown in Fig.~\ref{broadbandstorage}b. The recalled photons are shown as a light-blue trace in the histogram of coincidence detections in Fig.~\ref{broadbandstorage-2}. Analyzing the correlations between signal and recalled idler photons, we find $g_{si}^{(2)}=8.33 \pm 0.47$, which shows that the heralded photons after -- and hence also before -- storage, are indeed non-classical, 
and thus confirms the quantum nature of our light-matter interface. To improve the storage efficiency and thus the signal-to-noise ratio in the coincidence counts we increase the storage bandwidth of the memory to 16 GHz -- recall that the bandwidth of the input photons is 50 GHz. Due to restrictions imposed by the level structure of erbium (see Appendix Fig.~A2), the bandwidth increase is achieved by generating an additional 8 GHz-wide section separated from the first by around 20 GHz from edge to edge. This entails using two independent lasers operating at 1532.5 nm and 1532.7 nm wavelength for preparation of the two spectrally separated AFCs.
As expected, this leads to an improvement of the overall memory efficiency and thus coincidence count rate from $0.59 \pm 0.06$~Hz to $1.29 \pm 0.13$~Hz (see Fig.~\ref{broadbandstorage-2}), and an increase of $g_{si}^{(2)}$ to $18.2\pm0.9$. The number of AFCs, and thus the total storage bandwidth, can be further increased by employing additional pump lasers.

\begin{figure*}[t!]
\begin{center}
\includegraphics[width=\textwidth,angle=0]{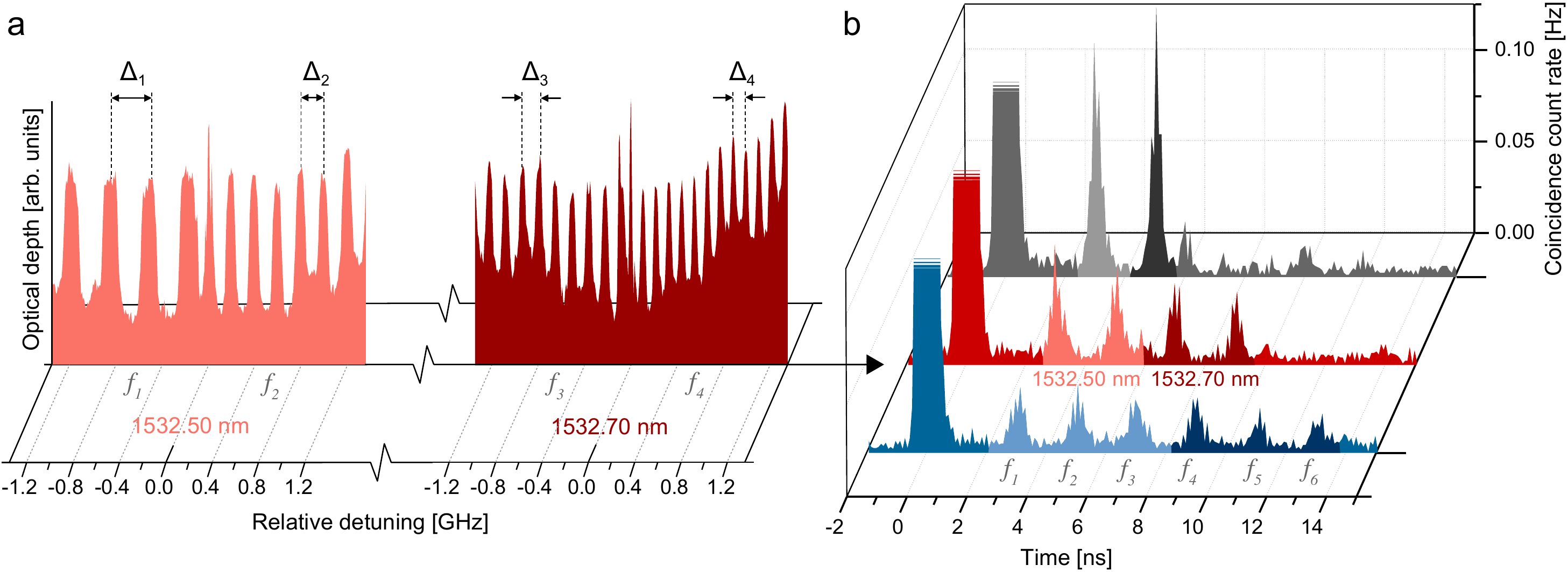}
\caption{\textbf{Multimode quantum storage.} \textbf{a) Creation of AFCs.} The total, currently addressable bandwidth of 18 GHz is divided into different spectral sections, each featuring an AFC with distinct peak spacing. Depicted is the case of four, 4.5 GHz-wide AFCs, created using two lasers operating at 1532.50 and 1532.70 nm. The AFCs feature peak spacings of 333 MHz, 200 MHz, 143 MHz and 111 MHz, corresponding to storage times of 3, 5, 7 and 9~ns, respectively. For each AFC, only a 1.3 GHz-wide section is shown.
\textbf{b) Storage and recall.}  When broadband heralded photons with 50 GHz bandwidth are mapped onto a two-section AFC, where each section extends over 9 GHz bandwidth and has a peak spacing of 333 MHz and 200 MHz, respectively, they are stored in two spectral modes and retrieved in two spectro-temporal modes, as shown in the back trace. Decreasing the bandwidth per AFC allows increasing the number of AFCs (spectral modes), as demonstrated with the storage of 4 and 6 spectral modes in the middle and the front trace, respectively. The modes are labeled $f_i$. For each mode, $g_{si}^{(2)}$ is measured to be larger than 2 (see Table \ref{g2numbers}).} 
\label{multimode}
\end{center}
\end{figure*}

Second, to demonstrate quantum storage with large time-bandwidth product we extend the memory storage time from 5 ns up to 50 ns. To this end we program a 16 GHz wide spectral region with two AFCs having tooth spacing ranging from 200 MHz to 20 MHz. For each case we map heralded photons onto the double AFC and collect coincidence statistics for the recalled photons as shown in Fig.~\ref{storagetime}. As further detailed in the Supplementary Information of \cite{saglamyurek2015a}, decoherence effects and imperfect preparation of the AFCs decreases the memory efficiency as the storage time increases. Nevertheless, as shown in the inset of Fig.~\ref{storagetime}, $g_{si}^{(2)}$ remains above 2 up to the maximal storage time of 50~ns, and thus we demonstrate the storage of non-classical light with a time-bandwidth product up to 800 (i.e. 16~GHz~$\times$~50~ns). This is an improvement over the current state-of-the-art by close to a factor of 40.

\begin{table}[t!]
	\begin{tabular}{ c | c || c | c| c}
		mode & storage &  \multicolumn{3}{ c }{$g_{si}^{(2)}$ for spectro-temporal modes:} \\[2pt]
		\cline{3-5}
		\hspace{1pt} \# \hspace{1pt} &time & \hspace{1pt} 9~GHz \hspace{1pt} & \hspace{1pt} 4.5~GHz & \hspace{1pt} 3~GHz \hspace{1pt}  \\[2pt]
		\hline
		\ $f_1$ & 3~ns \  & \ $16.8\pm 0.8$ \ & \ $8.0\pm 0.6$ \ & \ $5.7\pm 0.6$ \ \\[2pt]
		\ $f_2$ & 5~ns \  & \ $16.4\pm 0.8$ \ & \ $7.2\pm 0.6$ \ & \ $5.3\pm 0.6$ \ \\[2pt]
		\ $f_3$ & 7~ns \  &         --        & \ $6.3\pm 0.5$ \ & \ $5.4\pm 0.6$ \ \\[2pt]
		\ $f_4$ & 9~ns \  & 		--		  & \ $5.2\pm 0.5$ \ & \ $5.0\pm 0.6$ \ \\[2pt]
		\ $f_5$ & 11~ns \ &  		--		  &			--	  	 & \ $3.4\pm 0.5$ \ \\[2pt]
		\ $f_6$ & 13~ns \ &  		--		  &  		--		 & \ $3.4\pm 0.5$ \ \\[2pt]
	\end{tabular}
	\caption{Measured values of the cross correlation function of each recalled spectral mode $f_i$ for different AFC bandwidths and total numbers of modes. The storage times are given in column 2.}
	\label{g2numbers}
\end{table}\textbf{}

Third, to establish the multimode operation of our quantum light-matter interface, we divide the total currently accessible bandwidth -- 18~GHz for this experiment -- into different numbers of AFC sections, and program each section with a different storage time ranging from 3-13~ns. This allows identifying photons stored in different frequency modes (i.e. different AFC sections) through their recall time. In succession we perform measurements with: i) two AFCs of 9~GHz bandwidth; ii) four AFCs of 4.5~GHz bandwidth (this case is shown in Fig.~\ref{multimode}a); iii) six AFCs of 3~GHz bandwidth. The corresponding histograms of coincidence detections, depicted in Fig.~\ref{multimode}b, confirm that two, four and six modes, respectively, have been stored simultaneously. For each recalled spectro-temporal mode, we find that $g_{si}^{(2)}$, listed in Table \ref{g2numbers}, exceed the classical limit of 2, thereby confirming multimode storage of non-classical states of light in matter. 

\begin{figure}[t!]
\begin{center}
\includegraphics[width=\columnwidth,angle=0]{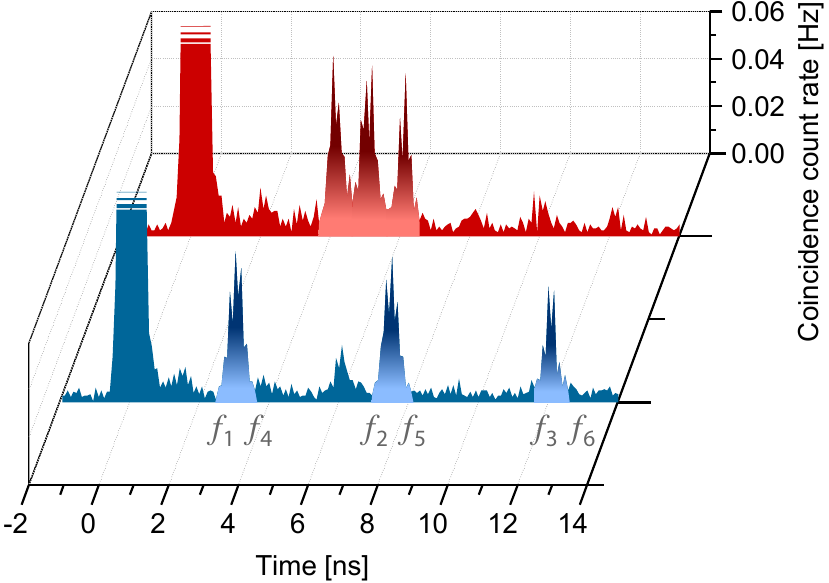}
\caption{\textbf{Pulse manipulation:} The AFC memory with total bandwidth of $2\times 8~\mathrm{GHz} =16$~GHz is tailored such that simultaneously absorbed photons in 6 spectral modes (from $f_{1}$ to $f_{6}$), each of 2.65~GHz bandwidth, are recalled in 3 temporal modes either spaced by 4.5~ns (front trace), or by 1~ns (back trace).
}
\label{manipulation}
\end{center}
\end{figure}

The frequency-to-time mapping described above was used to unambiguously distinguish between different frequency modes. This is convenient in our proof-of-principle demonstration, but may not be required in all applications. However, the ability to flexibly reconfigure our light-matter interface in terms of the number, bandwidth, storage time and frequency separation of the AFCs lends itself to versatile manipulation at the quantum level. In particular, it allows modifying the temporal shape of single photons by recalling different spectral modes at the same time and with adjustable phases. As an illustration Fig.~\ref{manipulation} shows two different mappings of 6 spectral modes onto 3 distinct temporal modes. We note that additional processing, e.g. pulse sequencing and temporal compressing, relying on the frequency-to-time mapping is possible by combining AFCs with frequency shifters, as demonstrated in \cite{saglamyurek2014a}. 

\subsection*{Discussion and Conclusion}

A central feature of our demonstration is the large multiplexed storage and manipulation capacity of our atomic interface for non-classical light. Yet, as noted above, there are clear avenues to increase this even further. Our additional investigations depicted in Appendix Fig.~A3 show that at least a 14~nm span of the erbium absorption line is suitable for quantum state storage. If we generate 8~GHz broad AFCs separated by 20 GHz, this yields a total of 64 AFCs covering a combined bandwidth of $64 \times 8~\mathrm{GHz} = 512$~GHz. Assuming 50~ns storage time one could thus reach a time-bandwidth product of more than $512~\mathrm{GHz} \times 50~\mathrm{ns} \approx 25000$, 
which is equal to the number of temporal--spectral modes that could be stored simultaneously.

On the other hand, a number of improvements are required for our multiplexed light-matter interface to be truly suitable for use in future quantum networks. 
First, imperfect optical pumping during the preparation of the AFCs currently results in a storage efficiency at 0.6 K of 1-2~\% (see the Supplementary Information of \cite{saglamyurek2015a}). However, as we describe in \cite{saglamyurek2015b}, we expect that it is possible to substantially improve the optical pumping, e.g. with lower temperatures and smaller erbium concentrations, and hence to approach unit efficiency under certain conditions.

Second, the $\sim$10~ms radiative lifetime of the $^4I_{15/2}\leftrightarrow^4I_{13/2}$ transition in erbium fundamentally limits the coherence time to 20 ms. However, coupling to so-called two-level systems \cite{macfarlane2006a}, which are intrinsic to disordered materials such as glass fibres, reduces the coherence time, and  the attainable storage time is hence currently restricted to a few tens of nanoseconds. While we anticipate that this value increases with smaller doping concentration, lower temperature \cite{staudt2006a, sun2006a} (see Supplementary Information of \cite{saglamyurek2015a} for further discussions), it is an open question whether or not it is possible to extend storage times to hundreds of $\mu s$, which would, e.g., allow building a quantum repeater based on spectral multiplexing \cite{sinclair2014a}. Yet, we note that not all applications of quantum memory require long storage times, and that the figure of merit for multiplexing schemes is the time-bandwidth product. Hence, our light-matter interface can be useful even without long storage times.  

Finally we point out that our storage device provides pre-programmed delays, which is sufficient for increasing the efficiency of multi-photon applications if one incorporates external elements that allow feed-forward-based mode-mapping, e.g. frequency shifters \cite{sinclair2014a}. However, for some applications, it may be desirable to perform the mode mapping during storage -- a familiar example being storage combined with read-out on demand, i.e. the possibility to select the recall time after the photon has been stored. One option to enable this feature is to utilize the Stark effect, which allows 'smearing' out the AFC peaks -- whose spacing determines the recall time -- after photon absorption \cite{lauritzen2011a}. This would  inhibit the rephasing of the collective excitation and hence reemission of light after $t_\mathrm{storage}=1/\Delta$, and lead to the possibility to recover the original photon a time $t_\mathrm{storage}=n/\Delta$ after absorption, where $n$ is an integer. Another possibility is to map the optical coherence -- created by the photon absorption -- onto a long-lived spin level (spin-wave storage) \cite{gundogan2015a,jobez2015a}, e.g. a hyperfine level (caused by the non-zero nuclear spin of $^{167}$Er$^{3+}$) \cite{hashimoto2011a,baldit2010a} or a superhyperfine level (caused by the interaction of Er with co-dopants in the host) \cite{thiel2010a}. However, this possibility remains challenging due to the complex and not fully characterized structure of these levels in erbium-doped fibres.  

Provided that the performance of our light-matter interface is improved as discussed, it can advance several quantum photonic applications. For instance, by programming AFCs with different storage times into different spectral regions, it can serve as a time-of-flight spectrometer, which would, for example, allow fine-grained, spectrally resolved photon measurements, including two-photon Bell-state measurements in a quantum repeater architecture based on frequency multiplexing \cite{sinclair2014a}. As a second example, our interface can be used to spectrally and temporally tailor single photon wave packets, as exemplified in our demonstrations, and thereby adapt their properties for subsequent interfacing with other quantum devices. More generally, by combining the interface with feed-forward-controlled frequency shifters, e.g. broadband electro-optic phase modulators, it can be turned into a programmable atomic processor for arbitrary manipulation of photonic quantum states encoded into time and/or frequency \cite{saglamyurek2014a}. This would find application in optical quantum computing in a single-spatial mode \cite{humphreys2013a} or in atomic-interface based photonic quantum state processing \cite{hosseini2009a, buchler2010a, nunn2013a, campbell2014a}. In addition, we note that the large multiplexing  capacity of our light-matter can be increased further by adapting multiplexing in spatial degree \cite{lan2009a}, and/or angular orbital momentum degree \cite{zhou2015a} with use of multimode erbium-doped fibers, as envisioned for future fiber-based classical and quantum networks \cite{bozinovic2013a}.   

In conclusion, we have presented a quantum light-matter interface that is suitable for multiplexed quantum-information-processing applications. More precisely, we have demonstrated a large time-bandwidth-product memory capable of multimode storage of non-classical light created via spontaneous parametric down-conversion, and showed that our light-matter interface can be employed for quantum manipulation of broadband heralded single photons. The fully pigtailed photon source and the fibre-based memory, along with telecommunication-wavelength operation and commercial availability, makes our demonstration important in view of building future fibre-based quantum networks.

\begin{center}
\textbf{METHODS}
\end{center}

\noindent
\subsection*{Erbium doped silica fibre}
We use a 20~m long, commercially available single-mode, 190~ppm-wt erbium-doped silica fibre specified to have 0.6~dB/m absorption at 1532~nm wavelength at room temperature. At 1~K we measure 0.1~dB/m absorption at 1532~nm wavelength (see Fig.~\ref{storagetime}). In addition to Er, the fibre is co-doped with P, Al and Ge. The fibre is spooled in layers around a copper cylinder with $\sim$4~cm diameter that is thermally contacted with the base plate of an adiabatic demagnetization refrigerator (ADR) maintained at about 0.6-0.8~K and exposed to a field of~$\sim$600~G inside a superconducting solenoid magnet. This setup induces about 70\% bending loss from input to output, mainly because of the insufficiently large diameter of the erbium-fibre spool. The fibre is fusion-spliced to standard single mode fibres (SMF-28) at each end with less than 5\% loss per splice.
 
\subsection*{Preparation of AFC} 
The memory is prepared using the setup described in Fig.~\ref{setup}. Frequency chirped laser pulses, applied 500 times per pumping cycle of 500~ms duration, allow frequency-selective optical pumping of erbium ions into a long-lived Zeeman level, and hence the formation of peaks and troughs of the AFC.
A polarization scrambler (PS) randomly changes the polarization of the pump light every 500~$\mu$s to ensure that heralded photons, which propagate counter to the optical pumping light, are absorbed regardless their polarization state \cite{jin2015b}. 
Upon completion of the optical pumping, a significant portion of the erbium ions accumulate in the auxiliary ground state Zeeman-level $\ket{s}$ (see Fig. \ref{broadbandstorage}), whose decay at 0.7~K and under a magnetic field of 600 Gauss is characterized by two lifetimes of 1.3~s and 26~s \cite{saglamyurek2015b}. 
However, a considerable amount of atoms remains in excited level $\ket{e}$, and subsequent spontaneous emission from these would mask all recalled photons. Thus, to eliminate this spontaneous emission noise we set a wait time of 300~ms, which is significantly larger than the 11~ms exited level lifetime.
Following the wait time, we store and retrieve many heralded single photons during a 700~ms measurement time. The retrieval efficiency of our AFC is about 1-2\%, which is primarily limited by residual absorption background as a result of incomplete population transfer during the optical pumping, (see Supplementary Information of \cite{saglamyurek2015a} for further details)

\subsection*{Superconducting nanowire single photon detectors}
The detection of the telecom-wavelength photons is carried out by a set of superconducting nano-wire single-photon detectors (SNSPDs) \cite{marsili2013a} attached to the base plate of the ADR and maintained at the same temperature as the memory. In our setup, the tungsten silicide (WSi) SNSPDs have efficiencies of about 70\% which includes the loss due to fibre-splices and bends. Their detection efficiencies exhibit a small polarization dependence of about 5\%. The time jitter of the detectors is around 250~ps, which allows us to resolve detection events separated by 1~ns (see Fig.~\ref{manipulation}). We measure the dark count rate of the detectors to be about 10~Hz, which results in a negligibly small contribution to accidental coincidences and hence high signal-to-noise ratios for the coincidence detections of the recalled photons.

\subsection*{Coincidence and $\mathbf{g^{(2)}}$ measurements}

All detection signals are directed to a time-to-digital converter (TDC), which allows recording detection times with a resolution of 80~ps. The detection of the heralding (\emph{idler}) photon on SNSPD1 is used to trigger (start) the TDC. The other member of the pair (the \textit{signal} photon) is detected by SNSPD2 -- either as recalled (delayed) photon after storage, or as a directly transmitted photon through the erbium fibre that features no extra delay -- and sent to the TDC, which records the time interval from the trigger signal. This allows us to generate histograms of the time-resolved coincidence detections between \textit{signal} and \textit{idler} photons (see, e.g., Fig.~\ref{broadbandstorage-2}).

To calculate $g_{si}^{(2)}$ using Eq.~\ref{g2}, we extract the values for the rates of the coincidence detection $R_{si}(t=0)$ and the individual detections $R_{s}$ and $R_{i}$ from coincidence histograms. Here we have explicitly stated the time difference between the coincidence detections to be 0. We note that, since the photon-pair generation process is spontaneous, there is no statistical correlation between subsequent pair generation events. Hence we can re-write the product $R_{s} R_{i}$  as $R_{si}(t\neq0)$, i.e. as the coincidence detection rates for \textit{signal} and \textit{idler} photons that are not members of the same photon pair, which is often referred to as accidental coincidence count-rate \cite{kuzmich2003a}. We extract $R_{si}(t=0)$ from a coincidence histogram by counting all detections within the ``coincidence peak" that is centred at time $t_0$ and has width $t_p$, and normalizing this number by measurement time and coincidence window width $t_p$. Similarly, $R_{si}(t\neq0)$ is evaluated by appropriate normalization of coincidence counts taken in a window of width $t_{bg}$ that is adjacent to the ``coincidence peak". Hence, to evaluate $g_{si}^{(2)}$ from experimental histograms of coincidence counts $R_{si}(t)$ we use
\begin{align}
	g_{si}^{(2)}=\frac{ t_{bg}\int_{t_0-t_p/2}^{t_0+t_p/2} R_{si}(t) dt}
		{t_p \int_{t_0+t_p/2}^{t_0+t_p/2+t_{bg}} R_{si}(t) dt}.
\end{align}
For all the cross-correlation values in this paper we used coincidence windows $t_p=t_{bg}=0.8$~ns.

\section*{Acknowledgments}
ES, MG, QZ, LG, DO and WT thank Jeongwan Jin, Neil Sinclair, Charles Thiel and Vladimir Kiselyov for discussions and technical support, and acknowledge funding through Alberta Innovates Technology Futures (AITF) and the National Science and Engineering Research Council of Canada (NSERC). Furthermore, W.T. acknowledges support as a Senior Fellow of the Canadian Institute for Advanced Research (CIFAR), and VBV and SWN partial funding for detector development from the DARPA Information in a Photon (InPho) program. Part of the research was carried out at the Jet Propulsion Laboratory, California Institute of Technology, under a contract with the National Aeronautics and Space Administration. LO and DN acknowledge Srico Inc. for their assistance with the fabrication of periodically poled lithium niobate wafers.

\section*{Additional information}

Correspondence and requests for materials should be addressed to W. Tittel (email: \mbox{wtittel@ucalgary.ca).}



\def\opone{\leavevmode\hbox{\small1\kern-3.8pt\normalsize1}}
\renewcommand{\figurename}{Appendix Figure}
\makeatletter
\newcommand{\manuallabel}[2]{\def\@currentlabel{#2}\label{#1}}
\renewcommand{\thefigure}{A\@arabic\c@figure} 
\renewcommand{\theequation}{A\arabic{equation}} 
\renewcommand{\thetable}{A\arabic{table}}
\makeatother

\begin{figure*}[h!]
\begin{center}
\includegraphics[width=0.60\textwidth,angle=0]{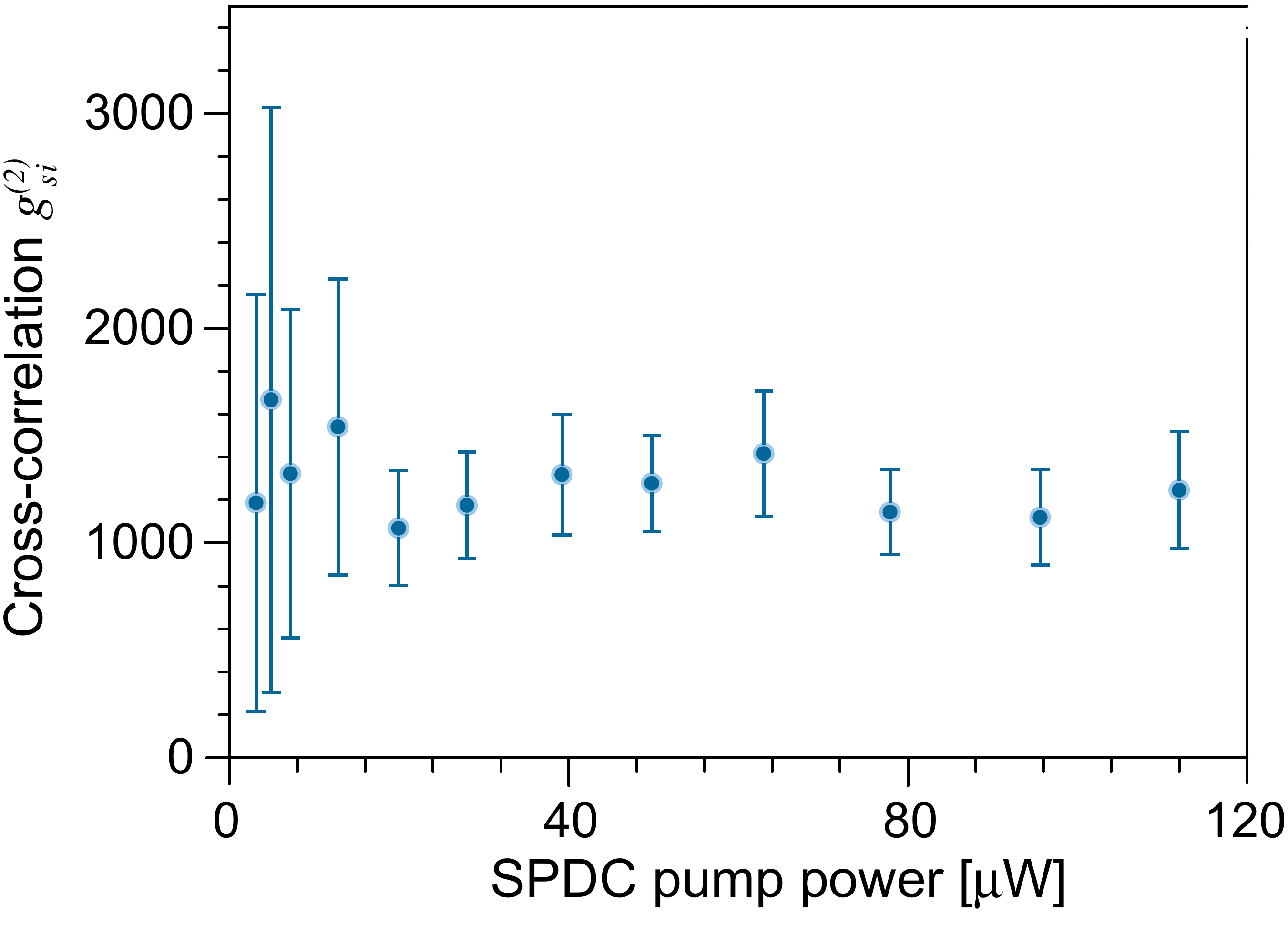}
\caption{\textbf{Characterisation of $g_{si}^{(2)}$ before storage vs. SPDC pump power.} We measure the second-order cross-correlation function given in Eq.~(\ref{g2}) as a function of SPDC pump power before storage in the AFC memory. All measured values of $g_{si}^{(2)}$ exceed 1000, which is substantially above the classical limit of 2 for thermal light fields -- having a photon number distribution following Bose-Einstein statistics). The large uncertainties of $g_{si}^{(2)}$ for low pump powers are due to small photon pair generation rates in conjunction with the relatively short data accumulating times.}
\label{power}
\end{center}
\end{figure*}

\begin{figure*}[h!]
\begin{center}
\includegraphics[width=1\textwidth,angle=0]{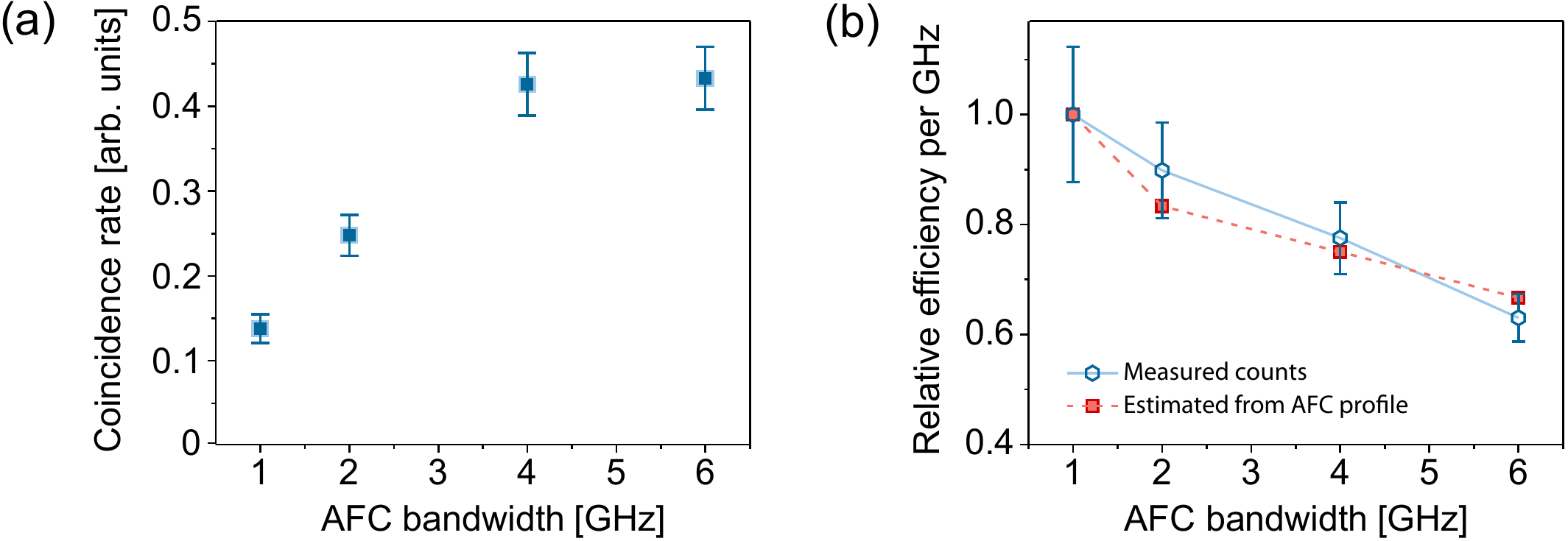}
\caption{\textbf{Storage efficiency vs. AFC bandwidth.} Due to the limited splitting and significant broadening of the Zeeman-levels -- employed as shelving levels for the population that is optically pumped to tailor an AFC -- the bandwidth of each individual AFC is limited. More precisely, as the AFC bandwidth increases, population removed from one edge of the AFC starts filling the troughs at its opposite edge, leading to a reduction of the recall efficiency. This behaviour can be seen from a), where the rate of detected heralded photons at the memory output is plotted as a function of the bandwidth of a single AFC at 1532~nm. As the bandwidth is increased the memory efficiency initially grows, but then reaches a plateaus. In b) we plot measured and estimated efficiencies per AFC bandwidth, normalized to 1 for a 1~GHz broad AFC. As expected, we find a monotonous decrease as the total bandwidth increases. 
}
\label{BW_echo}
\end{center}
\end{figure*}

\begin{figure*}[h!]
\begin{center}
\includegraphics[width=0.75\textwidth,angle=0]{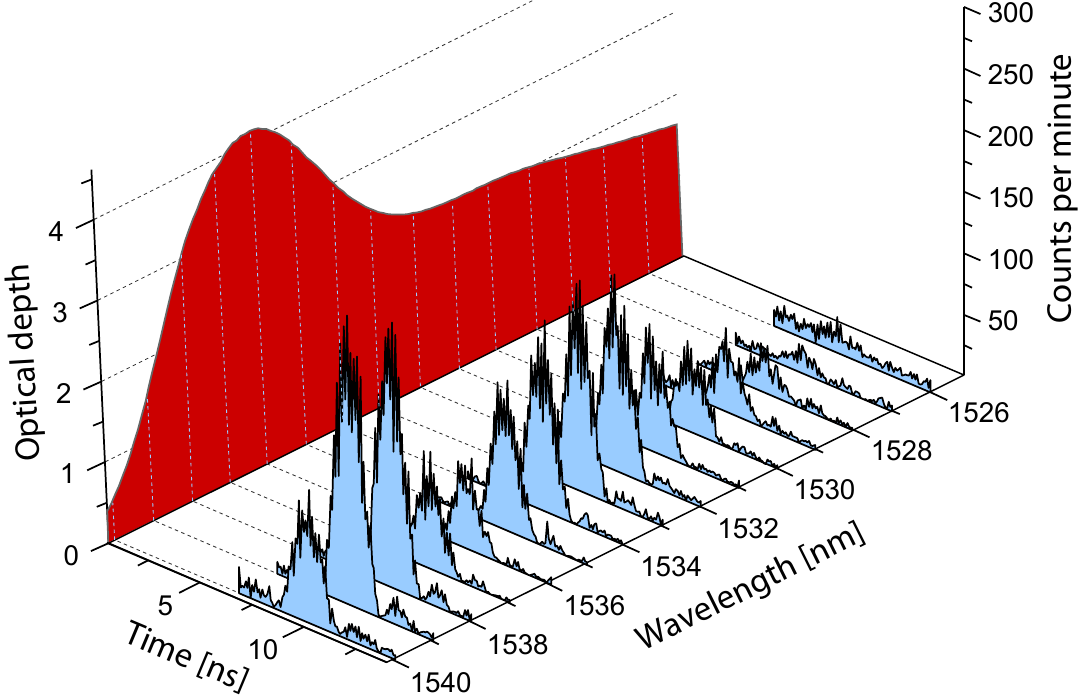}
\caption{\textbf{Total accessible AFC bandwidth.} To assess the total width of the inhomogeneously broadened absorption line of the erbium-doped fibre that is accessible for quantum storage, we tune the wavelength of the optical pumping laser to generate 1~GHz wide AFCs with 10~ns storage time over a spectral interval ranging from 1526~nm to 1540~nm wavelength. We then store 2~ns long attenuated laser pulses with mean photon number of $0.5\pm0.1$, generated at an effective mean rate of 100~kHz, in the AFC. We observe recalled photons in all cases, showing that the entire bandwidth of 14 nm is suitable for optical quantum memory (i.e. spin level lifetimes as well as optical coherence times are sufficiently large). The observed variation in the echo intensity is predominantly due to the different initial optical depths of the transition. The temperature of the fiber was 0.8~K and the magnetic field was 600~G.}
\label{totBW}
\end{center}
\end{figure*}

\FloatBarrier

\begin{centering}
	\section*{Appendix 1: INTEGRATED PHOTON PAIR SOURCE}
\end{centering}

In the following, we detail the structural properties and fabrication of our non-classical light source. Our heralded single photon source (HSPS) operates by creating photon pairs through spontaneous parametric down-conversion (SPDC) in a fibre-pigtailed PPLN crystal waveguide with type-0 phase matching.  Although the time at which photon pairs are generated by illumination with a CW pump laser is random and cannot be known precisely, the temporal correlations between a pair of photons created by SPDC are well-defined, and the detection of one photon can be exploited to successfully herald the arrival of its twin. In our HSPS, an optical fibre couples the pump photons that are generated from a CW laser into an optical waveguide, which is located inside the lithium niobate crystal.  The optical waveguide extends across the entire length of the 50~mm long, 0.5-mm thick Z-cut lithium niobate chip.  In the central region of the lithium niobate chip, the ferroelectric polarization of the lithium niobate is periodically inverted over a total length of approximately 40~mm.  The input pump photons are converted to photon pairs in this periodically poled region through the interaction of the pump field with this second order ($\chi^{(2)}$) nonlinear medium.  In order to generate photon pairs with high efficiency through SPDC, the periodically poled region and optical waveguides were designed to satisfy the following quasi-phase matching \cite{lim1989a,webjorn1989a} condition:
\begin{equation}
	\frac{n_p}{\lambda_p}-\frac{n_s}{\lambda_s}-\frac{n_i}{\lambda_i}=\frac{1}{\Lambda}
\end{equation}
where $\Lambda$ is the period of the poled domains and $n_p$ ($\lambda_p$), $n_s$ ($\lambda_s$), and $n_i$ ($\lambda_i$) are the effective refractive indices of the waveguide at the pump, signal, and idler wavelengths $\lambda_p$, $\lambda_s$, and $\lambda_i$, respectively.  The effective indices of refraction of the optical waveguides at the pump, signal, and idler wavelengths were determined by performing numerical analysis.  For generation of frequency degenerate photon pairs with $\lambda_{s(i)} \approx 1532$~nm from a pump photon with $\lambda_p \approx 766$~nm, the optimal period of the poled domains was observed to be approximately $17~\mu$m. 
 
The PPLN waveguides were fabricated using a Ti-indiffusion process, where a titanium layer was applied to the –Z surface of a lithium niobate crystal and Ti-indiffusion was achieved by heating the crystal in a furnace at $T \sim1000^{\circ}$C.  The PPLN waveguide width was $7~\mu$m, and single mode operation was observed at $\lambda \sim 1532$~nm for PPLN waveguide widths $< 8~\mu$m.  The photon pairs generated inside the PPLN waveguide are collected in an optical fibre, and SMF-28 fibre was used to ensure single mode propagation.   In order to achieve stable and reliable optical coupling, SMF-28 optical fibre connectors were attached to the PPLN crystal using UV-cured epoxy.  In order to reduce the optical coupling loss between the PPLN waveguide and the optical fibre and, therefore, increase the heralding efficiency of our single photon source, tapered and periodically segmented waveguide (PSW) geometries \cite{castaldini2007a}  were integrated into the optical waveguide design.  A detailed description of the optical modeling that supported the development of the PPLN waveguide and the PSW taper designs used in our HSPS has been presented in a prior publication, along with supporting optical test and characterization data \cite{oesterling2015a}. 

For this work, the CW pump power level was limited to less than $\sim100~\mu$W.  This power level constraint was a consequence of the photorefractive effect.  For pump power levels $< 100~\mu$W, the rate at which photon pairs are generated scales linearly with pump power.  At pump power levels $> 100~\mu$W the rate of photon pair generation no longer scales linearly with pump power, and the pair generation rate will start to saturate and eventually degrade as the power level is increased.  To enable operation at higher power levels, our team is currently designing Ti-indiffused PPLN waveguides to operate at $T > 120^{\circ}$C, because photorefractive effects are significantly reduced at elevated temperatures.  We are also designing, fabricating, and characterizing proton-exchanged and Zn-indiffused PPLN waveguides, which are less sensitive to photorefractive effects than Ti-indiffused waveguides.\\[14pt]


\begin{centering}
	\section*{Appendix 2: QUANTUM CORRELATIONS AND STORAGE IN AFC MEMORY}\label{corr}
\end{centering}

Below we discuss how the parameters that govern the storage of photons generated by the SPDC source determine the measured cross correlations both before and after storage. We start by restating Eq.~(\ref{g2}) in the main text for $g_{si}^{(2)}$ in terms of experimentally measurable quantities
\begin{equation}
 \label{g2_S2}
  g_{si}^{(2)}=\frac{R_{si}}{R_{acc}} \ ,
	\end{equation}
where $R_{si}$ is the photon pair coincidence count rate and $R_{acc}$ is the accidental count rate, which is identical to the quantity $R_{si}(t\neq 0)$ defined in the Methods section of the main text. $R_{acc}$ has in principle contributions from multi pair emissions of the source as well as detector dark counts. However, at the rate of 10~Hz, the latter is negligible. For the case in which the fibre memory is bypassed,  $R_{si}$ can be decomposed as
 \begin{equation}
 \label{Psi}
  R_{si}=0.5\eta_c^2\eta_d^2R+R_{acc} \ ,
	\end{equation}	
where $\eta_c$, $\eta_d$ and $R$ are the collection efficiency of signal and idler photon, the detection efficiency of the SNSPDs, and the photon pair generation rate, respectively. 
We extract $\eta_c$ from measurements of the coincidence and signal count rates, $R_{si}$ and $R_s$, respectively, and the detection efficiency of the SNSPD on the \textit{idler} side to be about 8.6\%. As shown in Fig.~\ref{power} we measure $g_{si}^{(2)}$ values in excess of 1000, which points to $R_{acc}$ being only a very small fraction of $R_{si}$.

With the addition of the AFC memory for the storage of one member of the photon pair, this expression can be written as 
 \begin{equation}
 \label{Psi_m}
  R_{si}^\prime=0.5\eta_c^2\eta_d^2\eta_sR+R_{acc}^\prime \ ,
	\end{equation}
where $\eta_s$  is the system efficiency for the recalled photon in a specific spectral and temporal mode, and $R_{acc}^\prime$ is the accidental coincidence rate measured when the AFC memory is operated. By inspecting Eqs.~\ref{g2_S2} - \ref{Psi_m}, we find that $g_{si}^{(2)}$ will be reduced if $R_{acc}^\prime > \eta_s R_{acc}$. Hence, $g_{si}^{(2)}$ will decrease either due to a low $\eta_s$ or due to additional noise induced by the memory. In the following we elaborate on both effects.

First, the system efficiency $\eta_s$ can be considered a product of three factors: the retrieval efficiency of the AFC memory $\eta_m$, the transmission factor $\eta_t$ that takes into account the optical loss in the erbium-doped fibre due to splices and bending as well as in the optical circulator, and a filtering factor $f$ ($0\leq f \leq 1$), which is given by ratio of the bandwidth of the AFC section of interest and the bandwidth of the input photons (50~GHz). While $\eta_t$, which we measured to be 14\%, is a constant in all experiments, both $\eta_m$ and $f$ depend on the preparation of the AFC. It is obvious that $f$ is directly proportional to the AFC bandwidth, and we also find that the recall efficiency $\eta_m$ decreases with increasing bandwidth of a single AFC, as described in Appendix Fig.~\ref{BW_echo}. Of course, a reduction of $\eta_s$ does not necessarily decrease $g_{si}^{(2)}$ unless a noise source that is independent of $\eta_s$ is present, e.g., ultimately,  detector dark counts.

Provided the transmission loss and the filtering factor are known, the AFC retrieval  efficiency  can  directly be determined for any spectro-temporal mode by time-resolved measurements of $R_{si}$ and $R_{si}^\prime$ (corresponding to the cases with and without memory, respectively), since $\eta_m$ is given by
\begin{equation}
	\eta_m=\frac{1}{\eta_t f}\frac{R_{si}^\prime-R_{acc}^\prime}{R_{si}-R_{acc}}.
	\end{equation}
This formula, combined with independently determined values for the filtering factor and $\eta_t$, is used to compute the memory efficiency $\eta_m$ plotted in Fig.~4 in the main text and Appendix Fig.~\ref{BW_echo}.

Next, we investigate the noise sources contributing to $R_{acc}^\prime$. We start be writing the accidental coincidence rate as
 \begin{equation}
 \label{noise}
  R_{acc}^\prime=\eta_s R_{acc}+R_{noise} \ ,
	\end{equation}
where $R_{acc}$ is the same as that defined in Eq.~(\ref{Psi}) --- note that it appears as a product with the system efficiency $\eta_s$. Hence, the cross-correlation function cannot be reduced due to any source of accidental coincidences that is already present without the memory. However, the term $R_{noise}$, which includes two sources that are unique to the operation of the memory, leads to a reduction of $g_{si}^{(2)}$. i) The first source of noise is the light spontaneously emitted from the small fraction of atoms that are excited during optical pumping and remain in the excited state after 300~ms of waiting time. In principle, this noise and the resulting accidental coincidences can be further suppressed by extending the waiting time and better optimization of the intensity of the optical pumping light. ii) The second source of noise is due to SPDC photons that are absorbed by atoms that do not contribute to the reversible mapping via the AFC i.e. atoms that either remain in a trough (and hence contribute to the AFC background), or that are outside of the AFC section. These photons are spontaneously re-emitted within the lifetime of the excited level (10~ms) and can be accidentally detected in coincidence with the heralded photons. 

To highlight the effect of spontaneous-emission noise, we rewrite Eq.~(\ref{g2_S2}) taking into account Eqs.~(\ref{Psi_m}) and (\ref{noise})
\begin{equation}
 \label{gsiagain}
  g_{si}^{(2)}=\frac{R_{si}^\prime}{R_{acc}^\prime}=\frac{0.5\eta_c^2\eta_d^2 R+ R_{acc}+R_{noise}/\eta_s}{R_{acc}+R_{noise}/\eta_s}.
	\end{equation}
This shows that even though the number of spontaneously emitted photons may be much smaller than the number of input SPDC photons within the AFC bandwidth, they can nevertheless cause an significant reduction in $g_{si}^{(2)}$. This is mainly a consequence of our small current system efficiency $\eta_s$, on the order of 0.1\%. 

There is a number of ways to reduce the effect of spontaneous emission noise, one obviously being to increase the AFC efficiency. Another path is to reduce the bandwidth mismatch between the input photons and the tailored AFC sections. In spectral multiplexing applications this could be accomplished by using multimode sources that produce photons only in spectral modes that are matched to the AFC sections. Another solution is to implement additional filtering to remove the spectrally unmatched portions of the input photons prior to the memory.\\[14 pt]



\begin{thebibliography}{10}
\expandafter\ifx\csname url\endcsname\relax
  \def\url#1{\texttt{#1}}\fi
\expandafter\ifx\csname urlprefix\endcsname\relax\def\urlprefix{URL }\fi
\providecommand{\bibinfo}[2]{#2}
\providecommand{\eprint}[2][]{\url{#2}}

\bibitem{england2011a}
\bibinfo{author}{England, D.~G.} \emph{et~al.}
\newblock \bibinfo{title}{High-fidelity polarization storage in a gigahertz
  bandwidth quantum memory}.
\newblock \emph{\bibinfo{journal}{J. Phys. B At. Mol. Opt. Phys.}}
  \textbf{\bibinfo{volume}{45}} (\bibinfo{year}{2012}).

\bibitem{michelberger2015a}
\bibinfo{author}{Michelberger, P.~S.} \emph{et~al.}
\newblock \bibinfo{title}{Interfacing ghz-bandwidth heralded single photons
  with a warm vapour raman memory}.
\newblock \emph{\bibinfo{journal}{New J. Phys.}} \textbf{\bibinfo{volume}{17}}
  (\bibinfo{year}{2015}).

\bibitem{england2015a}
\bibinfo{author}{England, D.~G.} \emph{et~al.}
\newblock \bibinfo{title}{Storage and retrieval of thz-bandwidth single photons
  using a room-temperature diamond quantum memory}.
\newblock \emph{\bibinfo{journal}{Phys. Rev. Lett.}}
  \textbf{\bibinfo{volume}{114}} (\bibinfo{year}{2015}).

\bibitem{ding2015a}
\bibinfo{author}{Ding, D.-S.} \emph{et~al.}
\newblock \bibinfo{title}{Raman quantum memory of photonic polarized
  entanglement}.
\newblock \emph{\bibinfo{journal}{Nat. Photon.}} \textbf{\bibinfo{volume}{9}},
  \bibinfo{pages}{332--338} (\bibinfo{year}{2015}).

\bibitem{nunn2008a}
\bibinfo{author}{Nunn, J.} \emph{et~al.}
\newblock \bibinfo{title}{Multimode memories in atomic ensembles}.
\newblock \emph{\bibinfo{journal}{Phys. Rev. Lett.}}
  \textbf{\bibinfo{volume}{101}} (\bibinfo{year}{2008}).

\bibitem{riedmatten2008a}
\bibinfo{author}{de~Riedmatten, H.}, \bibinfo{author}{Afzelius, M.},
  \bibinfo{author}{Staudt, M.~U.}, \bibinfo{author}{Simon, C.} \&
  \bibinfo{author}{Gisin, N.}
\newblock \bibinfo{title}{A solid-state light -matter interface at the
  single-photon level}.
\newblock \emph{\bibinfo{journal}{Nature}} \textbf{\bibinfo{volume}{456}},
  \bibinfo{pages}{773--777} (\bibinfo{year}{2008}).

\bibitem{afzelius2009a}
\bibinfo{author}{Afzelius, M.}, \bibinfo{author}{Simon, C.},
  \bibinfo{author}{de~Riedmatten, H.} \& \bibinfo{author}{Gisin, N.}
\newblock \bibinfo{title}{Multimode quantum memory based on atomic frequency
  combs}.
\newblock \emph{\bibinfo{journal}{Phys. Rev. A}} \textbf{\bibinfo{volume}{79}}
  (\bibinfo{year}{2009}).

\bibitem{usmani2010a}
\bibinfo{author}{Usmani, I.}, \bibinfo{author}{Afzelius, M.},
  \bibinfo{author}{de~Riedmatten, H.} \& \bibinfo{author}{Gisin, N.}
\newblock \bibinfo{title}{Mapping multiple photonic qubits into and out of one
  solid-state atomic ensemble}.
\newblock \emph{\bibinfo{journal}{Nat. Commun.}} \textbf{\bibinfo{volume}{1}}
  (\bibinfo{year}{2010}).

\bibitem{bonarota2011a}
\bibinfo{author}{Bonarota, M.}, \bibinfo{author}{Gou{\"e}t, J. L.~L.} \&
  \bibinfo{author}{Chaneli{\`e}re, T.}
\newblock \bibinfo{title}{Highly multimode storage in a crystal}.
\newblock \emph{\bibinfo{journal}{New. J. Phys.}} \textbf{\bibinfo{volume}{13}}
  (\bibinfo{year}{2011}).

\bibitem{gundogan2013a}
\bibinfo{author}{G{\"u}ndo{\u{g}}an, M.}, \bibinfo{author}{Mazzera, M.},
  \bibinfo{author}{Ledingham, P.~M.}, \bibinfo{author}{Cristiani, M.} \&
  \bibinfo{author}{de~Riedmatten, H.}
\newblock \bibinfo{title}{Coherent storage of temporally multimode light using
  a spin-wave atomic frequency comb memory}.
\newblock \emph{\bibinfo{journal}{New J. Phys.}} \textbf{\bibinfo{volume}{15}}
  (\bibinfo{year}{2013}).

\bibitem{sinclair2014a}
\bibinfo{author}{Sinclair, N.} \emph{et~al.}
\newblock \bibinfo{title}{Spectral multiplexing for scalable quantum photonics
  using an atomic frequency comb quantum memory and feed-forward control}.
\newblock \emph{\bibinfo{journal}{Phys. Rev. Lett.}}
  \textbf{\bibinfo{volume}{113}} (\bibinfo{year}{2014}).

\bibitem{tang2015a}
\bibinfo{author}{Tang, J.-S.} \emph{et~al.} 
\newblock \bibinfo{title}{Storage of multiple single-photon pulses emitted from a quantum dot in a solid-state quantum memory}.
\newblock \emph{\bibinfo{journal}{Nat. Commun.}} \textbf{\bibinfo{volume}{6}}
  \bibinfo{pages}{8652} (\bibinfo{year}{2015}).

\bibitem{jin2015b}
\bibinfo{author}{Jin, J.} \emph{et~al.}
\newblock \bibinfo{title}{Telecom-wavelength atomic quantum memory in optical
  fiber for heralded polarization qubits}.
\newblock \emph{\bibinfo{journal}{Phys. Rev. Lett.}}  (\bibinfo{year}{2015}).

\bibitem{saglamyurek2015a}
\bibinfo{author}{Saglamyurek, E.} \emph{et~al.}
\newblock \bibinfo{title}{Quantum storage of entangled telecom-wavelength
  photons in an erbium-doped optical fibre}.
\newblock \emph{\bibinfo{journal}{Nat. Photon.}} \textbf{\bibinfo{volume}{9}},
  \bibinfo{pages}{83--87} (\bibinfo{year}{2015}).

\bibitem{saglamyurek2015b}
\bibinfo{author}{Saglamyurek, E.} \emph{et~al.}
\newblock \bibinfo{title}{Efficient and long-lived zeeman-sublevel atomic
  population storage in an erbium-doped glass fiber}.
\newblock \emph{\bibinfo{journal}{Preprint at arXiv:1507.03012v1}}
  (\bibinfo{year}{2015}).

\bibitem{tapster1998a}
\bibinfo{author}{Tapster, P.~R.} \& \bibinfo{author}{Rarity, J.~G.}
\newblock \bibinfo{title}{Photon statistics of pulsed parametric light}.
\newblock \emph{\bibinfo{journal}{J. Mod. Opt.}} \textbf{\bibinfo{volume}{45}}
  (\bibinfo{year}{1998}).

\bibitem{bashkansky2014a}
\bibinfo{author}{Bashkansky, M.}, \bibinfo{author}{Vurgaftman, I.},
  \bibinfo{author}{Pipino, A. C.~R.} \& \bibinfo{author}{Reintjes, J.}
\newblock \bibinfo{title}{Significance of heralding in spontaneous parametric
  down-conversion}.
\newblock \emph{\bibinfo{journal}{Phys. Rev. A}} \textbf{\bibinfo{volume}{90}}
  (\bibinfo{year}{2014}).

\bibitem{saglamyurek2014a}
\bibinfo{author}{Saglamyurek, E.} \emph{et~al.}
\newblock \bibinfo{title}{An integrated processor for photonic quantum states
  using a broadband light-matter interface}.
\newblock \emph{\bibinfo{journal}{New J. Phys.}} \textbf{\bibinfo{volume}{16}}
  (\bibinfo{year}{2014}).

\bibitem{macfarlane2006a}
\bibinfo{author}{Macfarlane, R.~M.}, \bibinfo{author}{Sun, Y.},
  \bibinfo{author}{Sellin, P.~B.} \& \bibinfo{author}{Cone, R.~L.}
\newblock \bibinfo{title}{Optical decoherence in {E}r{$^{3+}$} doped silica
  fiber: Evidence for coupled spin-elastic tunneling systems}.
\newblock \emph{\bibinfo{journal}{Phys. Rev. Lett.}}
  \textbf{\bibinfo{volume}{96}} (\bibinfo{year}{2006}).

\bibitem{staudt2006a}
\bibinfo{author}{Staudt, M.~U.} \emph{et~al.}
\newblock \bibinfo{title}{Investigations of optical coherence properties in an
  erbium-doped silicate fiber for quantum state storage}.
\newblock \emph{\bibinfo{journal}{Opt. Commun.}}
  \textbf{\bibinfo{volume}{266}}, \bibinfo{pages}{720--726}
  (\bibinfo{year}{2006}).

\bibitem{sun2006a}
\bibinfo{author}{Sun, Y.}, \bibinfo{author}{Cone, R.~L.},
  \bibinfo{author}{Bigot, L.} \& \bibinfo{author}{Jacquier, B.}
\newblock \bibinfo{title}{Exceptionally narrow homogeneous linewidth in
  erbium-doped glasses}.
\newblock \emph{\bibinfo{journal}{Opt. Lett.}} \textbf{\bibinfo{volume}{31}},
  \bibinfo{pages}{3453--3455} (\bibinfo{year}{2006}).

\bibitem{lauritzen2011a}
\bibinfo{author}{Lauritzen, B.}, \bibinfo{author}{Minar, J.},
  \bibinfo{author}{de~Riedmatten, H.}, \bibinfo{author}{Afzelius, M.} \&
  \bibinfo{author}{Gisin, N.}
\newblock \bibinfo{title}{Approaches for a quantum memory at telecommunication
  wavelengths}.
\newblock \emph{\bibinfo{journal}{Phys. Rev. A}} \textbf{\bibinfo{volume}{83}}
  (\bibinfo{year}{2011}).

\bibitem{gundogan2015a}
\bibinfo{author}{G{\"u}ndo{\u{g}}an, M.}, \bibinfo{author}{Ledingham, P.~M.},
  \bibinfo{author}{Kutluer, K.}, \bibinfo{author}{Mazzera, M.} \&
  \bibinfo{author}{de~Riedmatten, H.}
\newblock \bibinfo{title}{Solid state spin-wave quantum memory for time-bin
  qubits}.
\newblock \emph{\bibinfo{journal}{Phys. Rev. Lett.}}
  \textbf{\bibinfo{volume}{114}} (\bibinfo{year}{2015}).

\bibitem{jobez2015a}
\bibinfo{author}{Jobez, P.} \emph{et~al.}
\newblock \bibinfo{title}{Coherent spin ccontrol at the quantum level in an
  ensemble-based optical memory}.
\newblock \emph{\bibinfo{journal}{Phys. Rev. Lett.}}
  \textbf{\bibinfo{volume}{114}} (\bibinfo{year}{2015}).

\bibitem{hashimoto2011a}
\bibinfo{author}{Hashimoto, D.} \& \bibinfo{author}{Shimizu, K.}
\newblock \bibinfo{title}{Population relaxation induced by the boson peak mode
  observed in optical hyperfine spectroscopy of ions doped in a silica glass
  fibre.}
\newblock \emph{\bibinfo{journal}{J. Opt. Soc. Am. B}}
  \textbf{\bibinfo{volume}{28}} (\bibinfo{year}{2011}).

\bibitem{baldit2010a}
\bibinfo{author}{Baldit, E.} \emph{et~al.}
\newblock \bibinfo{title}{Identification of {$\Lambda$}-like systems in
  ${E}r^{3+}$:{Y}$_2${S}i{O}$_5$ and observation of electromagnetically induced
  transparency}.
\newblock \emph{\bibinfo{journal}{Phys. Rev. B}} \textbf{\bibinfo{volume}{81}}
  (\bibinfo{year}{2010}).

\bibitem{thiel2010a}
\bibinfo{author}{Thiel, C.~W.} \emph{et~al.}
\newblock \bibinfo{title}{Optical decoherence and persistent spectral hole
  burning in {E}r{$^{3+}$}:{L}i{N}b{O}$_3$}.
\newblock \emph{\bibinfo{journal}{J. Lumin}} \textbf{\bibinfo{volume}{130}}
  (\bibinfo{year}{2010}).

\bibitem{humphreys2013a}
\bibinfo{author}{Humphrey, P.~C.} \emph{et~al.}
\newblock \bibinfo{title}{Linear optical quantum computing in a single spatial
  mode}.
\newblock \emph{\bibinfo{journal}{Phys. Rev.Lett.}}
  \textbf{\bibinfo{volume}{111}} (\bibinfo{year}{2013}).

\bibitem{hosseini2009a}
\bibinfo{author}{Hosseini, M.}, \bibinfo{author}{Sparkes, B.~M.},
  \bibinfo{author}{Longdell, G. H. J.~J.}, \bibinfo{author}{Lam, P.~K.} \&
  \bibinfo{author}{Buchler, B.~C.}
\newblock \bibinfo{title}{Coherent optical pulse sequencer for quantum
  application}.
\newblock \emph{\bibinfo{journal}{Nature}} \textbf{\bibinfo{volume}{461}}
  (\bibinfo{year}{2009}).

\bibitem{buchler2010a}
\bibinfo{author}{Buchler, B.~C.}, \bibinfo{author}{Hosseini, M.},
  \bibinfo{author}{Hetet, G.}, \bibinfo{author}{Sparkes, B.~M.} \&
  \bibinfo{author}{Lam, P.~K.}
\newblock \bibinfo{title}{Precision spectral manipulation of optical pulses
  using a coherent photon echo memory}.
\newblock \emph{\bibinfo{journal}{Opt. Lett.}} \textbf{\bibinfo{volume}{35}},
  \bibinfo{pages}{1091--1093} (\bibinfo{year}{2010}).

\bibitem{nunn2013a}
\bibinfo{author}{Nunn, J.} \emph{et~al.}
\newblock \bibinfo{title}{Enhancing multiphoton rates with quantum memories}.
\newblock \emph{\bibinfo{journal}{Phys. Rev. Lett.}}
  \textbf{\bibinfo{volume}{110}} (\bibinfo{year}{2013}).

\bibitem{campbell2014a}
\bibinfo{author}{Campbell, G.~T.} \emph{et~al.}
\newblock \bibinfo{title}{Configurable unitary transformations and linear logic
  gates using quantum memories}.
\newblock \emph{\bibinfo{journal}{Phys. Rev. Lett.}}
  \textbf{\bibinfo{volume}{113}} (\bibinfo{year}{2014}).

\bibitem{lan2009a}
\bibinfo{author}{Lan, S.-Y.} \emph{et~al.}
\newblock \bibinfo{title}{A multiplexed quantum memory}.
\newblock \emph{\bibinfo{journal}{Opt. Exp.}} \textbf{\bibinfo{volume}{17}},
  \bibinfo{pages}{13639} (\bibinfo{year}{2009}).

\bibitem{zhou2015a}
\bibinfo{author}{Zhou, Z.-Q.} \emph{et~al.}
\newblock \bibinfo{title}{Quantum storage of three-dimensional
  orbital-angular-momentum entanglement in a crystal}.
\newblock \emph{\bibinfo{journal}{Phys. Rev. Lett.}}
  \textbf{\bibinfo{volume}{115}} (\bibinfo{year}{2015}).

\bibitem{bozinovic2013a}
\bibinfo{author}{Bozinovic, N.} \emph{et~al.}
\newblock \bibinfo{title}{Terabit-scale orbital angular momentum mode division
  multiplexing in fibers}.
\newblock \emph{\bibinfo{journal}{Science}} \textbf{\bibinfo{volume}{340}},
  \bibinfo{pages}{1545--1548} (\bibinfo{year}{2013}).

\bibitem{marsili2013a}
\bibinfo{author}{Marsili, F.} \emph{et~al.}
\newblock \bibinfo{title}{Detecting single infrared photons with 93\% system
  efficiency}.
\newblock \emph{\bibinfo{journal}{Nature Photon.}}
  \textbf{\bibinfo{volume}{7}}, \bibinfo{pages}{210--214}
  (\bibinfo{year}{2013}).

\bibitem{kuzmich2003a}
\bibinfo{author}{Kuzmich, A.} \emph{et~al.}
\newblock \bibinfo{title}{Generation of nonclassical photon pairs for scalable
  quantum communication with atomic ensembles}.
\newblock \emph{\bibinfo{journal}{Nature}} \textbf{\bibinfo{volume}{423}},
  \bibinfo{pages}{731--734} (\bibinfo{year}{2003}).

\expandafter\ifx\csname url\endcsname\relax
  \def\url#1{\texttt{#1}}\fi
\expandafter\ifx\csname urlprefix\endcsname\relax\def\urlprefix{URL }\fi
\providecommand{\bibinfo}[2]{#2}
\providecommand{\eprint}[2][]{\url{#2}}

\bibitem{lim1989a}
\bibinfo{author}{Lim, E.}, \bibinfo{author}{Fejer, M.}, \& \bibinfo{author}{Byer, R.}
\newblock \bibinfo{title}{Efficient quasi-phase-matched blue second-harmonic generation in LiNbO$_3$ channel waveguides by a second-order grating}.
\newblock \emph{\bibinfo{journal}{Electron. Lett.}} \textbf{\bibinfo{volume}{25}}
  \bibinfo{pages}{174} (\bibinfo{year}{1989}).
%
\bibitem{webjorn1989a}
\bibinfo{author}{Webjorn, J.}, \bibinfo{author}{Laurell, F.}, \& \bibinfo{author}{Arvidsson, G.}
\newblock \bibinfo{title}{Domain inversion in MgO-diffused LiNbO$_3$}.
\newblock \emph{\bibinfo{journal}{J. Lightwave Technol.}} \textbf{\bibinfo{volume}{7}}
  \bibinfo{pages}{1597} (\bibinfo{year}{1989}).
%
\bibitem{castaldini2007a}
\bibinfo{author}{Castaldini, D.}, \bibinfo{author}{Bassi, P.}, \bibinfo{author}{Tascu, S.}, \bibinfo{author}{Aschieri, P.}, \bibinfo{author}{De Micheli, M.P.}, \& \bibinfo{author}{Baldi, P.}
\newblock \bibinfo{title}{Soft-Proton-Exchange Tapers for Low Insertion-Loss LiNbO$_3$ Devices}.
\newblock \emph{\bibinfo{journal}{J. Lightwave Technol.}} \textbf{\bibinfo{volume}{25}}
  \bibinfo{pages}{1588–1593} (\bibinfo{year}{2007}).
%
\bibitem{oesterling2015a}
\bibinfo{author}{Oesterling, L.}, \bibinfo{author}{Monteiro, F.}, \bibinfo{author}{Krupa, S.}, \bibinfo{author}{Nippa, D.}, \bibinfo{author}{Wolterman, R.}, \bibinfo{author}{Hayford, D.}, \bibinfo{author}{Stinaff, E.}, \bibinfo{author}{Sanguinetti, B.}, \bibinfo{author}{Zbinden, H.}, \& \bibinfo{author}{Thew, R.}
\newblock \bibinfo{title}{Development of a Photon Pair Source using Periodically Poled Lithium Niobate and Fiber Optic Components}.
\newblock \emph{\bibinfo{journal}{Journal of Modern Optics}} \textbf{\bibinfo{volume}{62}}
  \bibinfo{pages}{1-10} (\bibinfo{year}{2015}).

\end{thebibliography}
\end{document}